\documentclass[preprint2]{aastex}

\begin{document}
\title{The mass-to-light ratio of rich star clusters}

\shorttitle{M/L of rich star clusters}
\shortauthors{Boily, Fleck, Lan\c{c}on, Renaud}

\author{Christian M. Boily\altaffilmark{}, Jean-Julien Fleck\altaffilmark{}, Ariane Lan\c{c}on and Florent Renaud}
\affil{Observatoire astronomique \& CNRS UMR7550, Universit\'e de Strasbourg\\11 rue de l'Universit\'e, Strasbourg F-67000}
\email{boily@astro.u-strasbg.fr}

\begin{abstract}
We point out a strong time-evolution of the mass-to-light conversion factor $\eta$ commonly used to estimate masses of unresolved star clusters from observed cluster spectro-photometric measures. We present a series of gas-dynamical models coupled with the Cambridge stellar evolution tracks to compute line-of-sight velocity dispersions and half-light radii weighted by the luminosity. We explore a range of initial conditions, varying in turn the cluster mass and/or density, and the stellar population's IMF. We find that $\eta$, and hence the estimated cluster mass, may increase by factors as large as 3 over time-scales of $50$ million years. We apply these results to an hypothetic cluster mass distribution function (d.f.) and show that the d.f. shape may be strongly affected at the low-mass end by this effect. Fitting truncated isothermal (Michie-King) models to the projected light profile leads to over-estimates of the concentration parameter $c$ of $\delta c\approx 0.3$ compared to the same functional fit applied to the projected mass density.
\end{abstract}


\section{Introduction}
The formation of star clusters in bursts of star formation during galactic mergers has attracted much attention since the ground-breaking study by Schweizer (1986). Young clusters can account for $\sim 20\% $ of the UV light flux of their sample starburst galaxies (compared with $< 1\%$ for the Milky Way; e.g. Meurer et al. 1995). Such numbers bring to focus the role that star formation in clusters plays in shaping the overall (galactic) stellar mass function. Proto-typical cases where cluster formation has been a spectacular manifestation of interaction-induced starbursts are the merging systems NGC 4038/39 (the Antenn\ae) and the nearby galaxy M82 (recent interaction with M81). Many of the brightest clusters in these galaxies have estimated ages on the order of $10^7$ years, based on optical and near-IR spectra. High resolution spectroscopic data and HST images have been used to measure velocity dispersions and estimate virial masses. The line-of-sight velocity dispersion $\sigma_\mathrm{1d}$ (LOSVD) relates to the mass $M$ and projected half-light radius $r_\mathrm{hp}$ of a cluster in virial equilibrium through
\begin{equation} 
M = \eta \frac{r_\mathrm{hp}\sigma_\mathrm{1d}^2}{G} \ , \label{eq:eta}
\end{equation}
where $\eta$ is a dimensionless free parameter. A number of authors have set $\eta \approx 10 $ in their studies to derive $M$ from (\ref{eq:eta}) (Sternberg 1998; Mengel et al. 2002; Smith \& Gallagher 2001; McCrady et al. 2003; Maraston et al. 2004). McCrady et al. give a derivation of this value of $\eta$. Mengel et al. (2002) quote a range from $5.6 - 9.7$ for King (1966) models with concentration parameter in the range 0.5 to 2.5, which corresponds to most galactic globular clusters. We discuss the value of $\eta$ further below. Dynamical masses derived from (\ref{eq:eta}) have been compared to the stellar masses of synthetic populations, using both standard (field) and non-standard stellar initial mass functions (henceforth IMF; see Kroupa 2002 for a review). The data were found to be inconsistent with a universal IMF (Mengel et al. 2002; Smith \& Gallagher 2001). In particular, several clusters in M82 were found to be over-luminous with respect to their estimated mass (low mass-to-light ratio, $M/L_\ast$). This suggests that M82 clusters may form with a top-heavy stellar IMF (Smith \& Gallagher 2001; McCrady et al. 2003).

\begin{figure}[t]
\begin{center}
\includegraphics[width=8cm,bb = 25 10 600 600,clip]{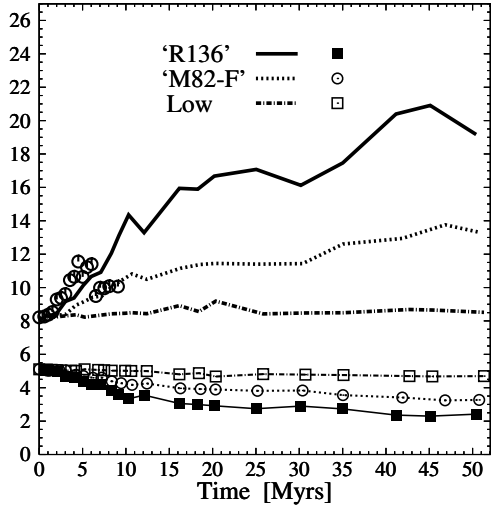}
\caption{Parameter $\eta$ (top lines) and half-light radius $r_\mathrm{hp}$ (bottom lines with symbols) versus time. The initial models were unsegregated King $\Psi/\sigma^2 = 6$ models. Results are shown for models with different initial surface density (cf. text). When the mass range is increased from 17 $\mathrm{M_{\odot}}$ to 50 $\mathrm{M_{\odot}}$, $\eta$ rises sharply over the first 10 Myr (open circles, case `M82-F'); thereafter it rejoins the curve displayed.}
\label{fig:etavstime}
\end{center}
\end{figure}

\section{Setup}
Much of what we report in this section draws from Boily et al. (2005).

\subsection{Numerical method}
The equations of motion were integrated numerically based on the gas-dynamical approach pioneered by Larson (1970) and developed further by Louis \& Spurzem (1991) to include anisotropic velocity fields. The method leans on an analogy between exchange of kinetic energy through star-star interactions and the classical heat-diffusion process of fluid dynamics (Lynden-Bell \& Eggleton 1980; Heggie \& Ramamani 1988). The implementation in spherical coordinates we used is largely due to Louis \& Spurzem (1991) and Giersz \& Spurzem (1994) but extended to include a spectrum of stellar masses (Spurzem \& Takahashi 1995).

The mass spectrum is sampled at constant logarithmic increments in the interval $\{m_0,m_1\}$; we used 14 bins in our standard runs ($\delta\ln m / M_{\odot} \approx 0.329$). To each mass bin corresponds the same initial (continuous) radial density profile. Star-star interactions (including those between stars of the same mass bin) lead to the diffusion of kinetic energy. Roughly speaking, the resulting change in the velocity dispersion of each mass bin causes a readjustment of the density profile. This is obtained using a semi-implicit Henyey integration. The gravitational potential is then updated by applying Poisson's equation (see Louis \& Spurzem 1991).

We checked that the mass-segregation time obtained from the models is in quantitative agreement with N-body and Fokker-Planck integrators by evolving Plummer models with three species of stars and following constant-mass Lagrangian radii in time for all three components (cf. Spitzer \& Shull 1975, Fig.~1). Spurzem \& Takahashi (1995) report excellent agreement from their own two-component test calculations.

\subsection{Stellar evolution} 
The different mass components are evolved according to the Cambridge tracks (Pols et al. 1998; Hurley et al. 2000). The tracks are efficiently coded in the form of fitting functions of the kind first presented by Eggleton, Tout \& Fitchett (1989). The functions return the current bolometric luminosity, radius, mass and metal abundance for given time and initial metal abundance $Z_0$; we set $Z_0 = 0.02$ (solar abundance) throughout.

A filter can be applied to the bolometric luminosity from model stellar atmospheres, and the total flux in a specified waveband read by interpolation polynomials from the Basel stellar library (Lejeune et al. 1998). We have mapped the stellar luminosity near the strongest near-IR CO bandhead ($\lambda \simeq 2.2 - 2.29\ \mathrm{\mu m}$) and the CaII triplet ($\lambda \simeq 8200 - 8600\ \mathrm{\AA}$) together with the bolometric limit (all wavelengths). Since our results are not strongly dependent on waveband issues, we will discuss only quantities computed from bolometric fluxes.

\subsection{Models} 
All models were scaled to $N = 500,000$ member stars and a total mass of $M \simeq 2 \times 10^5 \ \mathrm{M_\odot}$ when an Salpeter mass function is implemented. This is significantly less than what is measured for massive clusters in starburst galaxies, where membership may well exceed $10^6$. This point is important because the time-scale for segregation scales with $N/\ln N$. However we may explore the importance of varying the dynamical evolution time by constructing models with different sizes and central densities, while keeping the total mass, and $N$, constant. This approach is attractive because two systems with the same mean density will have the same dynamical time $t_\mathrm{cr}$, independently of their mass and size. We first setup three models with identical $\Psi/\sigma^2 = 6$ King parameter but each with different half-light radius. The model labeled `M82-F' has a mean surface density $\approx 1.5\times 10^4\ \mathrm{M_\odot/pc^2}$ or half the value we derive for that cluster from the data of Smith \& Gallagher (2001). The densest model is labeled `R136', in reference to the 30 Doradus cluster (central volume density $\sim 10^6 \ \mathrm{M_\odot/pc^3}$; Brandl et al. 1996). Finally, a third low-density model was evolved for comparison (labeled `Low').

\section{Results \& Applications}
\subsection{Reference models}
We start with the densest model `R136' for which we compute the shortest relaxation time $t_\mathrm{rh}$. Since $M,\ r_\mathrm{hp} $ and $\langle\sigma^2_\lambda\rangle$ are all known from the simulations, we solve for $\eta$ directly from (\ref{eq:eta}). The run of $\eta$ in time is displayed on Fig.~\ref{fig:etavstime}. The rapid increase of $\eta$ from an initial value $\approx 8.2$ is striking. After 15 Myr of evolution, $\eta$ has more than doubled. The projected half-light radius $r_\mathrm{hp}$ is displayed alongside $\eta$. The radius $r_\mathrm{hp}$ decreases steadily in time, a direct result of the migration of massive stars toward the centre. Note the slow but systematic rise of $\eta$ after $\approx 35$ Myr, when $\eta > 20$ (factor $>$ 2.5 from its initial value. The results are robust and can be extended to other stellar IMF (Fleck et al. 2006).

The time-evolution of $\eta$ for all three models is displayed on Fig.~1. Both $\eta$ and $r_\mathrm{hp}$ for the `Low' model stay essentially constant throughout. However the case `M82-F' (initial surface density $> 10 \ \mathrm{M_\sun/pc^2}$) shows unmistakable evolution, suggesting that for clusters of such or higher mean surface density we may no longer presume a time-independent mass-to-light parameter $\eta$. This has to be interpreted with care since unresolved, bright clusters will be more massive than the ones modelled here; despite this caveat, evolution will always take place in the central regions where the density peaks. This has immediate bearing on fitting cluster light profiles.

The LOSVD changes relatively little over time in comparison: we measure a monotonic decrease of $\langle\sigma_{\lambda}\rangle$ from $\approx 10.4$ to 9.3 km/s (or, -10.6\%) for the system as a whole, although for individual components evolution was more significant: down $\approx 30\%$ for the most massive stars, while the lightest component enjoys an increase of a comparable magnitude. These effects can all be traced back to the dynamical mass segregation.

\subsection{Embedding in a merger simulation} 
In this section we ask what effect the background galactic potential may have on the formation and early evolution of a star cluster. More specifically, we ask whether the tidal field is an ingredient that matters when it comes to understanding cluster formation in mergers. The results reported here are taken from a doctoral dissertation by Fleck (2007). 

We built two equal mass bulge/disk/halo galaxy models to simulate an Antenn\ae{}-like major merger with initial conditions similar to Barnes's (1988). We set up self-gravitating models using the {\tt magalie} software included with the Nemo\footnote{http://bima.astro.umd.edu/nemo/} stellar dynamical package (version 3.2.4; Teuben 1995). The equations of motion were integrated with gyrfalcON (Dehnen 2002) using a Plummer softening kernel with $\varepsilon=0.01$ in units where the gravitational constant $G=1$ and a numerical unit of time translates to 25 Myr. The total mass of each galaxy is equal to $7$ mass units, where the exponential disk scale-length and mass $r_\mathrm{d} = M_\mathrm{d} = 1$. The bulge is an Hernquist (1990) model with a scale parameter $a=1$ and total mass $M_\mathrm{b}=1$. Thus our progenitor galaxies are ``S0'' galaxies with relatively warm Toomre $Q=1.5$ disk parameter to ensure local stability to fragmentation modes. Note that our model galaxies differ mostly from Barnes's in that he used a lighter King (1966) model for the bulge. The dark matter halo was generated from a truncated isothermal distribution as described in Hernquist (1993). The mass ratio between individual particles $\simeq 1$ to minimize two-body heating; a total of $700,000$ particles were used for each galaxy, for a mass resolution of $(1.15 - 2.85) \times 10^5 \ \mathrm{M_{\odot}}$ for the different components. Converted to physical units, the simulations matched well the kinematics of the Antenn\ae{} (peak systemic velocity of 130 km/s) but for a total baryonic mass of $\simeq 9.2\times10^{10} \ \mathrm{M_{\odot}}$, some twice that measured for that system (cf. Hibbard et al. 2001).

The galaxies were set on a prograde encounter with tidal tails developing on either side of the orbital plane, implying non-coplanar disks. Fig.~V.17 of Fleck (2007, p.154) shows the orbit of the centers of mass of the two galaxies together with the orbit of the associated point-mass problem initially on a bound but eccentric orbit. The extent and mass of the dark matter halos are such that the galaxies are braking after the first passage ($t\approx11$) before falling back together and merging. Tidal tails form soon after the first passage and continue to expand until the end of the simulation at $t\approx 30$ (for a graphical time-sequence, see Fleck 2007, Fig.~IV.9, p.123).

\begin{figure*}[t]
\begin{center} 
\includegraphics[width=\linewidth,bb = 0 390 600 740,clip]{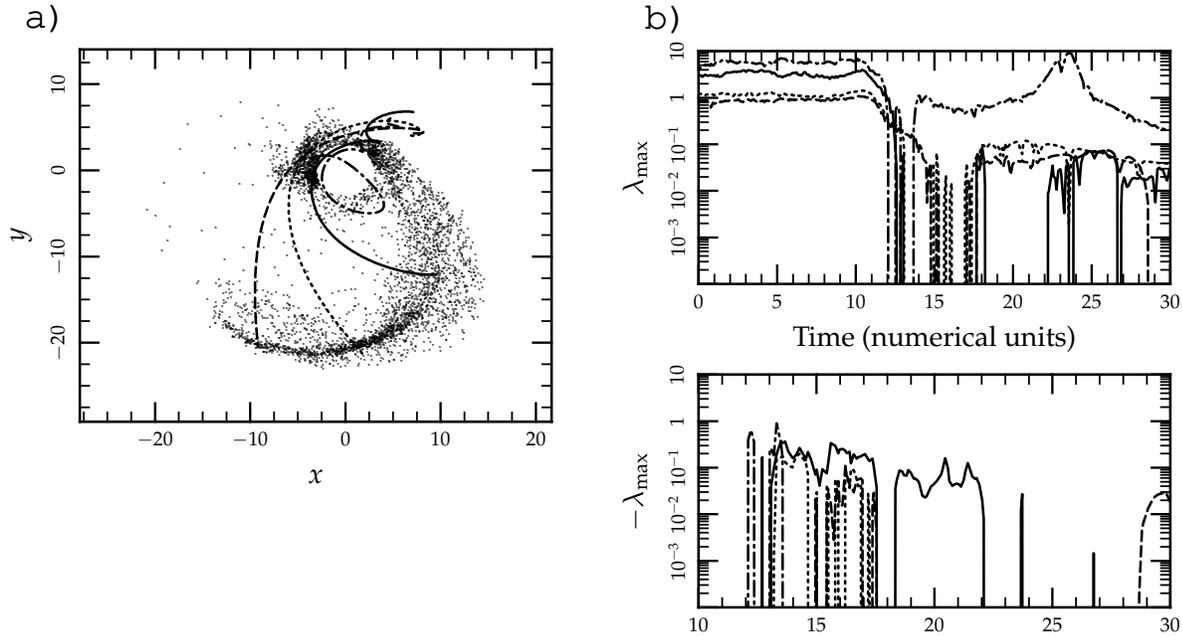}
\caption{a) Left-hand panel: Four orbits up to the final simulation time. All particles inside a radius $r < 3$ have been erased to avoid overcrowding. b) Right-hand panels: Time-evolution of the maximum eigenvalue measured along each orbit. Note the logarithmic scale. The top panel shows positive values, the lower panel negative values of compressive tides. }
\label{fig:individual:orbit}
\end{center}
\end{figure*}

\subsection{Individual orbits}
Our mass resolution of $\approx 2\times 10^5\ \mathrm{M_{\odot}}$ implies that individual disk and bulge particles represent cluster-size star forming regions, which we follow on their orbit. To explore a full range of formation histories, we chose several orbits that ended up either in the tails or remained close of the nuclear region at the end of the simulation. We retrieved the tidal tensor at the particle position using a second order finite difference scheme from the forces at six points distributed on a box of radius $\simeq \varepsilon$ ($\varepsilon = 40$ pc). The forces were directly summed over all the particles. We then computed the eigenvalues $\{\lambda_i\}$ of the tidal tensor using the fortran routines of Kopp (2008) and looked for regions that undergo purely compressive modes $(\max\{\lambda_i\} <0)$. On figure~\ref{fig:individual:orbit}, we graph a selection of orbits and their maximum eigenvalue as a function of time. Note that the orbit going back into the nuclei (dot-dashed line) experiences the strongest tidal force especially near the second passage, at and around time $t\approx 23$. Nevertheless, this orbit also experienced a fully compressive mode for a cumulative time interval of one unit. The orbit which spends the longest total period in a compressive mode ($\delta t \approx $ 8 time units, or $\sim 100$ Myr, solid line) is the one staying closest to the nuclei after a short trek in the tidal arms. The other two orbits also experience compressive modes immediately after the first passage, between $t=12$ and $t=18$.

Fig.~1 of Mengel et al. (2005) shows that young massive clusters are dispatched throughout the region surrounding the tidal bridge linking the two nuclei. To determine whether the distribution in space of young clusters may coincide with the compressive modes in the current-day configuration, we isolated a series of bodies that ended in the overlap region of the merger simulation and found that in several instances the bodies were still embedded in a volume of compressive tidal forces. Altogether we find that the fraction of bodies experiencing a compressive tidal field during the simulation rises from $\simeq 2\%$ in the initial configurations at rest, to $\simeq 15\%$ during both the first and second passage. It is clear from the statistics that tidal fields help push in and hold together the material that soon will burst to form a large ensemble of clusters (Renaud et al. 2008). This gathering of mass has long been evoked as a possible mechanism of dwarf galaxy formation with gas dissipation (e.g., Bournaud et al. 2006). Here we see that the same mechanism is operative at the very heart of a galaxy merger on scales that match those of rich star formation regions.

\acknowledgments
CMB is grateful to the organizers for an invitation to present the results of this study.

\end{document}